\documentclass[prl,aps,showpacs,twocolumn]{revtex4}
\usepackage{epsf}
\usepackage[dvips]{color}
\newcommand {\be}{\begin{equation}}
\newcommand {\ee}{\end{equation}}
\newcommand {\bea}{\begin{eqnarray}}
\newcommand {\eea}{\end{eqnarray}}

\begin{document}

\title{Spontaneous soliton formation and modulational instability
in Bose-Einstein condensates}
\author{L.~D. Carr}
\affiliation{Laboratoire Kastler Brossel, Ecole Normale
Sup\'erieure, 24 rue Lhomond, 75231 Paris, France}
\author{J. Brand}
\affiliation{Max Planck Institute for the Physics of Complex
Systems, N\"othnitzer Stra{\ss}e 38, 01187 Dresden, Germany\\}

\date{\today}

\begin{abstract}

The dynamics of an elongated attractive Bose-Einstein condensate
in an axisymmetric harmonic trap is studied.  It is shown that
density fringes caused by self-interference of the condensate
order parameter seed modulational instability.  The latter has
novel features in contradistinction to the usual homogeneous case
known from nonlinear fiber optics. Several open questions in the
interpretation of the recent creation of the first matter-wave
bright soliton train [Strecker {\it et al.} Nature {\bf 417} 150
(2002)] are addressed.  It is shown that primary transverse
collapse, followed by secondary collapse induced by
soliton--soliton interactions, produce bursts of hot atoms at
different time scales.
\end{abstract}


\maketitle

Solitons are ubiquitous in nature, appearing in systems as diverse as
DNA, shallow water, and laser light pulses in nonlinear fiber
optics~\cite{agrawal1}. Modulational instability (MI), an effect well
known in the latter context, is the process by which a constant-wave
background becomes unstable to sinusoidal modulations due to the
presence of a focusing nonlinearity, leading to a pulsed wave, called
a bright soliton train~\cite{hasegawa2}.  Recently, the first
matter-wave bright solitons were created from attractive Bose-Einstein
condensates (BEC's) in elongated harmonic
traps~\cite{carr29,strecker1}.  In the latter experiment a train of
from four to ten solitons was formed; however, unlike the solitons
resulting from MI of a uniform initial wavefunction in a constant
potential in one dimension~\cite{hasegawa2}, these solitons were
apparently stabilized by repulsive soliton--soliton interactions.
Moreover, of the $\sim 3\times 10^5$ initial atoms present when the
focusing nonlinearity was turned on, only $\sim 10\%$ survived to form
the soliton final state, implying massive collapse. It was
suggested~\cite{strecker1,khawaja2002} that MI,
in this new context, creates such a repulsively interacting soliton
train.

In the following, the MI of a {\it
non-uniform} initial state in the presence of a harmonic potential
is studied both analytically and numerically in the context of the
mean field of the BEC.  The conditions by which the transverse
dimensions may bring about {\it primary collapse}, either of the
entire condensate in the radial direction, or of individual
solitons as they are formed by MI, as well
as {\it secondary collapse} induced by soliton--soliton
interactions, are detailed. As a matter-wave bright soliton train
is in fact a train of self-contained BEC's easily guided and
manipulated by electromagnetic fields, it is important to
understand how it is made and how it may be kept stable. Moreover,
bright solitons, which have already revolutionized the
communications industry~\cite{agrawal1}, are, in the context of
the BEC, predicted to have applications in atom interferometry and
quantum frequency standards~\cite{rozhdestvenskii2002}.

The results of our analysis differ strongly from those
of previous studies~\cite{hasegawa2,khawaja2002}.  Firstly, it is
shown that self-interference of the order parameter seeds MI
at different times and with a steadily decreasing
wavelength.  This is true for any non-uniform initial density with the
exception of a Gaussian.  Secondly, it is shown that the ensuing
solitons may have any relative phase, and may therefore interact both
repulsively and attractively.  Thirdly, it is shown that both primary
and secondary collapse occurs.
This leads to the testable experimental
prediction of bursts of hot atoms emitted at
different time scales and to the conjecture that trajectories of
remaining solitons are stabilized by selection, as detailed below.

The 3D Nonlinear Schr\"odinger equation (NLS) or Gross-Pitaevskii
equation which describes the mean field of the BEC is written
as~\cite{dalfovo1}
\begin{equation}
\label{eqn:gpe3d} [ -\hbar^2 \nabla^2/2m
+g\,N\,|\Psi|^2 +V(\vec{r}) ]\Psi =i\hbar\partial_t\Psi \, ,
\end{equation}
where $V(\vec{r})\equiv m(\omega_{\rho}^2\rho^2
+\omega_z^2 z^2)/2$, $g\equiv 4\pi\hbar^2 a_s/m$, $a_s$ is the $s$-wave
scattering length,
$m$ is the atomic mass, $N$ is the number of condensed atoms, the
condensate order parameter $\Psi=\Psi(\vec{r},t)$ has been normalized to one, and
axisymmetric confinement has been assumed. Note that for negative
scattering length, or attractive nonlinearity, solutions are
liable to collapse in certain parameter regimes, as shall be
discussed below. In the case of strongly anisotropic confinement,
one may adiabatically separate the slow longitudinal from the fast
transverse degrees of freedom. The adiabatically varying
transverse state obeys a 2D NLS which shows an instability towards
collapse. The criterion for stability found by numerical
integration of the 2D NLS is \be\label{eqn:transv} \eta \equiv
-8 \pi a_s N |\psi(z,t)|^2 < 11.7 \, , \ee where $N |\psi(z,t)|^2$ is
the local axial line density of the
condensate~\cite{weinstein1983}.
If adiabaticity is violated, collapse can also happen at weaker
nonlinearity due to transverse oscillations on a time scale
$\pi/\omega_{\rho}$~\cite{pitaevskii1996}. When the longitudinal
dynamics is significantly slower than this time scale, the
adiabatic separation is valid. If, additionally, the transverse
nonlinearity is weak, {\it i.e.}, $|\eta| \ll 1$, the longitudinal
equation reduces to the quasi-1D NLS
\be [-\hbar^2\partial_z^2/2m
+g_{{\rm 1D}} N \left|\psi\right|^2 + m\omega_z^2 z^2/2
]\psi=i\hbar\,\partial_t\,\psi \,
,\label{eqn:gpe1d}\ee
where $g_{{\rm 1D}}\equiv 2\,a_s\,\omega_{\rho}\hbar$ is the
quasi-1D coupling constant~\cite{carr30}, provided
$l_{\rho}\gg |a|$~\cite{olshanii1}, with
$l_{\rho}\equiv\sqrt{\hbar/(m\omega_{\rho})}$.

In order to understand the mechanism of MI
for a non-uniform initial density profile and in the presence of a
non-constant potential, it is necessary to briefly review
MI in the uniform case, which is well known
from fiber optics~\cite{hasegawa2}. A linear response analysis
reveals that, for attractive nonlinearity, a small sinusoidal
modulation of a uniform state $\psi_0$ with wavenumber $k$ grows
exponentially at a rate $\gamma$ given by \be
\gamma^2=-\frac{\hbar^2}{4m^2} \left[k^2-\frac{2 N m
|g_{\mathrm{1D}}| \, |\psi_0|^2}{\hbar^2}\right]^2
+\frac{|\psi_0|^4 N^2 |g_{\mathrm{1D}}|^2 }{ \hbar^2} \, . \,
\label{eqn:mis1} \ee The maximum growth rate
$\gamma_{\mathrm{mg}}= 2 \omega_{\rho}\, |a_s|\,  N\,|\psi_0|^2$
is obtained at wavenumber $k_{\mathrm{mg}}=1/ \xi$, where $\xi =
l_{\rho}/\sqrt{4 |a_s|\, N\,|\psi_0|^2}$ is the effective 1D
healing length of the condensate~\cite{akh1992}. Note that
$N|\psi_0|^2$ is the line density of the condensate. Growth occurs
only if $\gamma^2>0$, which implies $0<k<k_{\mathrm{max}}=\sqrt{2}
k_{\mathrm{mg}}$. This means that nonlinear focusing can only be
seeded by modulations of sufficiently long wavelength and is
fastest at the length scale of $2\pi \xi$.

\begin{figure}[t]
\begin{center}
\epsfxsize=7.8cm \leavevmode \epsfbox{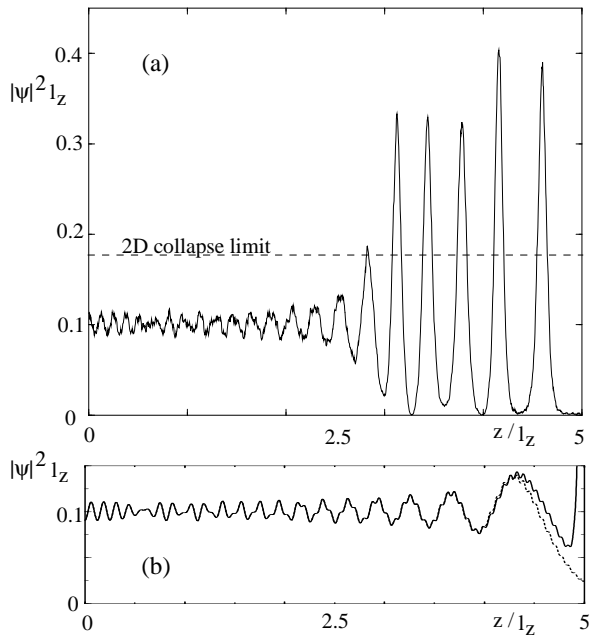}
\caption{\label{fig:1} Shown is the right half of the axial density of
a BEC in a harmonic potential, as determined by (a) numerical solution
of the {\it nonlinear} Schr\"odinger equation, and (b) analytical
solution of the {\it linear} Schr\"odinger equation (dashed: exact,
solid: approximate), at $t= 0.016\times 2\pi/\omega_{\rho}$, after 1D
evolution from an initial box state of length $L/l_z=10$ centered at
the origin. The formation of solitons first at the edge of the
condensate in (a) results from the longer wavelength of the linear
fringes, as seen in (b), since MI takes place at a wavelength $\sim
2\pi\xi= 0.43 l_z$ in this
simulation, using the parameters of Strecker {\it et
al.}~\cite{strecker1,foot1}.}
\end{center}
\end{figure}

\begin{figure}[t]
\begin{center}
\epsfxsize=7.8cm \leavevmode \epsfbox{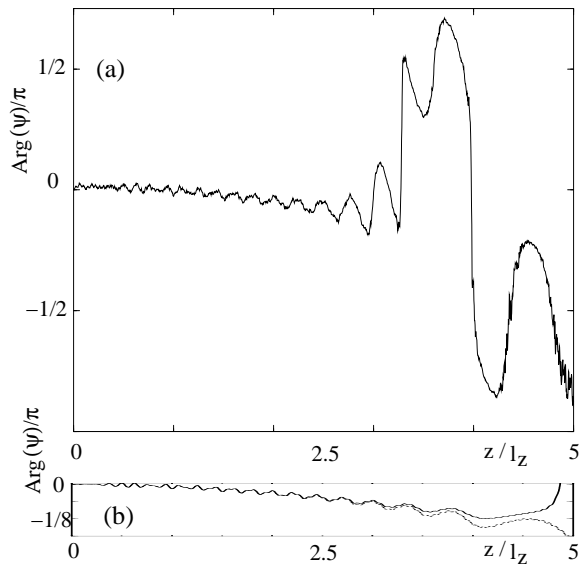}
\caption{\label{fig:2} Shown is the phase of the order parameter
corresponding to Fig.~\ref{fig:1}(a) and (b). The phase difference
$\Delta\phi$ between solitons may be seen by comparing to the
large density peaks of Fig~\ref{fig:1}(a): $\Delta\phi$ is highly
sensitive to the nonlinearity, and may take on arbitrary values
between $0$ and $2\pi$, in contrast to the usual case of
MI in fiber optics, where it is restricted
to $0$. At later stages in the simulation the relative phases
drift as the solitons decouple. }
\end{center}
\end{figure}

For {\it non-uniform} initial density profiles, self-interference
fringes in the order parameter can provide the necessary seed. To
understand this phenomenon, it is instructive to first consider
the development of the wavefunction for the ideal gas ($a_s=0$).
As a model, we shall consider the case of an initial state
$\psi(z,0)$ which is a rectangular function of height $\psi_0$ and
width $L$ centered at the origin. Recall that the Feynman
propagator which determines the evolution of the wavefunction in
the {\it linear} Schr\"odinger equation is defined by
$\psi(z,t)=\int dz' \,G(z,t;z',0) \psi(z',0)$. In the case of a
harmonic potential, $G$ may be determined exactly by
semi-classical methods to be~\cite{storey1}~\footnote{Up to a
global phase $\nu = \nu(t)$, called the Maslov index, which has no
physical significance for the observables in the problem at hand,
since the important quantity for soliton interactions is relative
phase.},
\be G = \frac{\exp \left\{{i}( z^2 - {2zz'}/{\cos \tau} +z'^2) /
({2 l_{z}^2 \tan \tau}) \right\}} {l_{z}\sqrt{2\pi i|\sin \tau|}}
\, , \, \label{eqn:fp2} \ee
where $\tau = \omega_z t$ and $l_z\equiv \sqrt{\hbar/(m\omega_z)}$ is
the axial harmonic
oscillator length.  This gives
%
%
an analytic expression for the evolution of the order parameter, 
which may be more easily understood in the limit
$\omega_z t \ll 1$:
\bea \frac{|\psi|^2}{|\psi_0|^2} \simeq 1+
\sqrt{\frac{8l_z^2\omega_z
t}{\pi}}\left[\frac{\sin(k_{+}z+\delta-\frac{\pi}{4})}{L+2z}\right.\nonumber\\
+\left.\frac{\sin(k_{-}z+\delta-\frac{\pi}{4})}{L-2z}\right]
+\frac{4 l_z^2\omega_z t}{\pi}
\left\{\frac{L^2+4z^2}{(L+2z)^2(L-2z)^2}\right.\nonumber\\
+\left.\frac{\cos[(k_{+}-k_{-}) z]}{(L+2z)(L-2z)}\right\}
+\frac{1}{2}\omega_z^2 t^2+\mathcal{O}[(\omega_z t)^{5/2}]\,
,\,\label{eqn:fp4} \eea
\bea \theta\simeq \sqrt{\frac{2l_z^2\omega_z
t}{\pi}}\left[\frac{\sin(k_{+}z+\delta+\frac{\pi}{4})}{L+2z}
+\frac{\sin(k_{-}z+\delta+\frac{\pi}{4})}{L-2z}\right]\nonumber\\
 +\frac{l_z^2 \omega_z t}{\pi}
\left\{-\frac{z^2\pi}{l_{z}^4}+\frac{\cos[2(k_{+}z+\delta)]}{(L+2z)^2}
+\frac{\cos[2(k_{-}z+\delta)]}{(L-2z)^2}\right.\nonumber\\
+\left.\frac{\cos[(k_{+}+k_{-})z+2\delta]}{(L+2z)(L-2z)}\right\}
+\mathcal{O}[(\omega_z t)^{3/2}]\, ,\,\label{eqn:fp5} \eea
\be k_{\pm}\equiv \frac{\sec(\omega_z t)z \pm
L}{2l_{z}^2\sin(\omega_z t)}\, ,\, \delta\equiv\frac{L^2
\cot(\omega_z t)}{8 l_{z}^2}\, ,\,|z|<\frac{L}{2}\,
,\,\label{eqn:fp6} \ee
where $\theta\equiv\mathrm{Arg}(\psi)$.  Figures~\ref{fig:1}(b)
and \ref{fig:2}(b) show the density and phase structure given by
Eqs.~(\ref{eqn:fp4}-\ref{eqn:fp6}), respectively. Note that the
above expansions converge very slowly as $|z|\rightarrow L/2$
where the initial wavefunction is dicontinuous.  

The above expressions for the evolution of the phase 
and density of the wavefunction may be understood as follows.  All {\it
trigonometric} terms represent quantum self-interference and
produce fringes which seed MI.  The
wavenumber $k_{\pm}$ is dependent on both time and position; the
wavelength is longer and the amplitude of oscillations higher at
the edges of the box than in the center, and the overall
wavelength increases as a function of time.  As these interference
terms are to linear order independent of $\omega_z$ (since
$l_z^2\omega_z=\hbar/m$), they must result solely from the
non-uniform initial density profile. One may also understand this
result by observing that Eq.~(\ref{eqn:fp2}) is, to order
$\omega_z t$, identical with the free space Feynman propagator in
one dimension.  In contrast, the {\it non-trigonometric} terms are
caused by the harmonic potential: in Eq.~(\ref{eqn:fp4}) they lead
to a monotonic increase in the mean density, where the order
$(\omega_z t)^2$ term is independent of position; and in
Eq.~(\ref{eqn:fp5}) they develop the overall phase profile
harmonically.  The development of the density and phase of the
wavefunction for the geometry of Ref.~\cite{strecker1} of $L\sim
10 \,l_z$ is shown in Figs.~\ref{fig:1} and~\ref{fig:2}.

%
%
%
%

%
\begin{figure}[t]
\begin{center}
\epsfxsize=7.8cm \leavevmode  \epsfbox{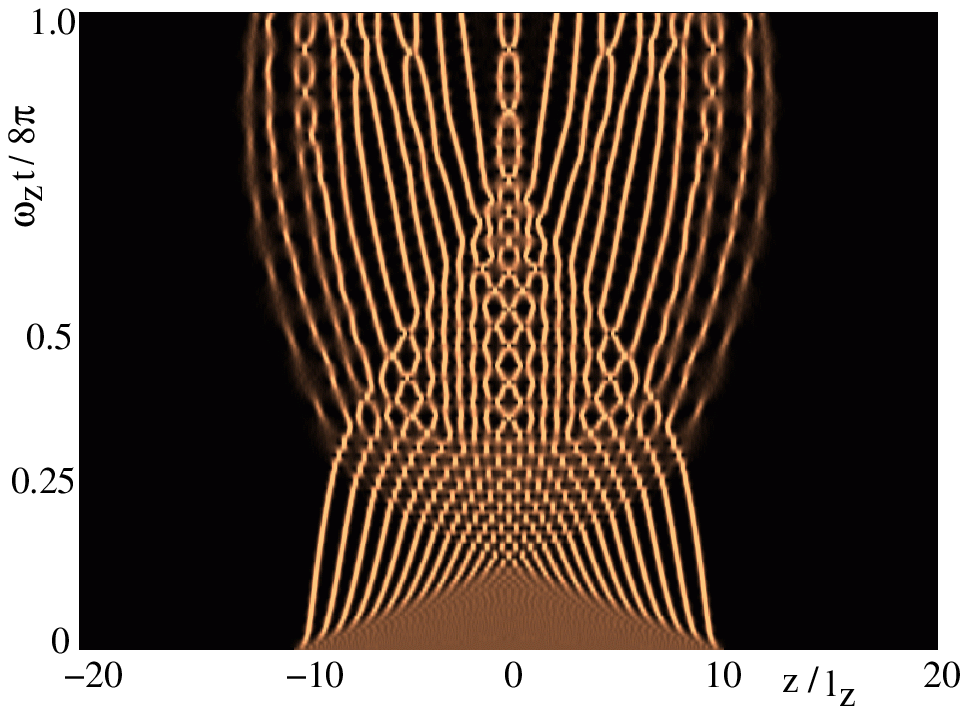}
\caption{\label{fig:3}  Shown is the 1D evolution of the axial
density of an elongated Bose-Einstein condensate when the
interactions are suddenly tuned negative, starting from a box-like
initial state. Here the parameters of the Strecker
experiment~\cite{strecker1} are used. At early times, small
fringes develop as the time-slice of Fig.~\ref{fig:1} clearly
shows; once the MI is seeded, the
nonlinearity focuses the fringes to produce solitons, the leading
edge of which is seen at the bottom of the figure as a wide
triangle shape. Lines visible above this edge are solitons. Their
complicated interaction pattern clearly shows many examples of
both attractive and repulsive soliton interactions. }
\end{center}
\end{figure}
%

MI of a non-uniform initial state, both with
and without an additional harmonic potential, may be understood in
light of the above considerations.  In Fig.~\ref{fig:3} is shown
the numerical evolution of the axial line density $N|\psi|^2$ in
Eq.~(\ref{eqn:gpe1d}) for the parameters of Strecker {\it et
al.}~\cite{strecker1} and our model of an initial square function.
The fringes produced by self-interference of the condensate order
parameter $\psi$ attain the critical wavelength for MI
of $\sqrt{2}\pi\xi$ first at the edges of the cloud,
as shown in Figs.~\ref{fig:1} and~\ref{fig:2}.  In
Fig.~\ref{fig:3} one notes that the solitons are therefore formed
last in the center of the trap.  In this latter stage they are
also formed closer together, as, due both to the harmonic
potential (see Eq.~(\ref{eqn:fp5})) and the global focusing effect
of the nonlinearity, the axial density has increased, and
$\xi\propto |\psi|^{-1}$.  Note that although the early phase structure in
Fig~\ref{fig:2} depends only weakly on the harmonic potential, in later stages of evolution the trap affects the phase strongly.

One may ask if the seed fringes are caused by the sharpness of the square
function edges in our model.  In 
the particular case $\omega_z t=\pi/4$, Eq.~(\ref{eqn:fp2}) takes the 
form of a Fourier transform, 
$\psi(z,t)\propto\int dz'\exp(-i\,zz'/l_z^2)  \psi(z',0)$.  
Thus it is immediately apparent that all initial wavefunctions 
with the exception of a Gaussian must develop fringes.  In order 
to test the full time development for an experimentally 
relevant initial density, a Thomas-Fermi
profile of harmonic trapping~\cite{dalfovo1} with $\psi\propto
\sqrt{1- (2z/L)^2}$ for $|z|<L/2$ and $\psi=0$ for $|z|\geq L/2$
was also studied and was found to have qualitatively the same
evolution in the linear Schr\"odinger equation, {\it i.e.},
fringes of longer wavelength towards the outer edges of the
density profile and an overall increase in wavelength with time.
The nonlinear evolution also shows the same qualitative behavior
as Fig.~\ref{fig:3}, with solitons forming spontaneously first
near the edges and later toward the center.

Besides the initial linear density fringes and nonlinear
MI, the condensate can undergo primary
transverse collapse on a time scale $\pi/\omega_{\rho}$, according
to Eq.~(\ref{eqn:transv}).
In the Strecker experiment, a rough estimate assuming a constant
initial density yields $\eta \approx 6.3$. In
fact, $\eta$ will exceed this value locally
due to the slanted initial profile produced by creating the BEC on
one side of the trap~\cite{strecker1} and transverse collapse is
likely to occur on the higher density side of the profile before
MI develops.  However, even for a uniform
initial state, the critical value of $\eta$ is reached during
soliton formation as indicated in Fig.~(\ref{fig:2}). Note that,
the center of mass of the BEC, created at the side of the trap,
trivially oscillates with the harmonic oscillator frequency and
decouples completely~\cite{garciaripoll2} from the relative
motion, the latter of which is analyzed in this work.

After the soliton train has spontaneously formed, even if
transverse collapse is avoided, individual solitons can undergo
primary 3D collapse, which dominates over transverse
instabilities. The static condition for three-dimensional soliton
stability in the absence of longitudinal confinement is given
numerically by~\cite{carr30}
\be N_{\mathrm{crit}}|a_s|/l_{\rho}= 0.627\ldots \, . \,
\label{eqn:3dcol}\ee
Dynamical effects, as for example breathing of the soliton, can lower
this condition further. In light of a recent experiment on a single
matter-wave bright soliton~\cite{carr29}, it is anticipated that such
a three dimensional collapse will eject the excess number of atoms,
leaving behind a solitonic core which is near the limit of
Eq.~(\ref{eqn:3dcol}).  In the simulation of
Figs.~\ref{fig:1}-\ref{fig:3}, the 26 solitons formed carry an average
of $\approx 1.9 N_{\mathrm{crit}}$ atoms where $N_{\mathrm{crit}}
\approx 6100$ for the parameters of the Strecker experiment. 
These
results are supported by a coarse estimate based on
Eq. (\ref{eqn:mis1}), which predicts formation of 22 solitons, and by
simulations of the 3D NLS analogous to the one of
Figs.~\ref{fig:1}-\ref{fig:3}, which shows primary collapse
immediately after formation of the first solitons (not shown).

Finally, the coherent overlap of solitons, not themselves subject
to two or three dimensional collapse, can cause secondary
collapse. The dynamics of binary soliton interactions are
determined by their relative phase $\Delta\phi$ and amplitude
$\Delta A$~\cite{gordon1}. For $-\pi/2<\Delta\phi<\pi/2$ they
attract each other and exchange mass. Consequently, they can
undergo collapse if, during their interaction, they temporarily
violate Eq.~(\ref{eqn:3dcol}). For $\pi/2<\Delta\phi<3\pi/2$ and
$\Delta A=0$ the interaction is repulsive and there is no mass
exchange; the solitons are stable against secondary collapse
induced by interactions. However, for $\Delta A\neq 0$ the
relative phase cycles periodically which, if it crosses over into
the regime $-\pi/2<\Delta\phi<\pi/2$, can again cause collapse.
Based on the above and the experiment of Ref.~\cite{carr29}, we
conjecture that, in the Strecker experiment, collisions of
solitons close to $N_{\mathrm{crit}}$ led, by several stages of
secondary collapse, to a final configuration of a small number of
solitons with stable trajectories.  We emphasize that these
results differ from those found in Ref.~\cite{khawaja2002}, where
it was suggested that the phase difference 
$\Delta\phi$ between adjacent solitons originates from
  quantum fluctuations and is restricted to values close to $\pi$,
  thus stabilizing their trajectories. We have shown that MI
  of a nonuniform initial state may produce arbitrary
  $\Delta\phi$ already in a mean-field model. We conjecture that
it is the secondary collapse processes which
determine a stable trajectory of solitons in the final state with
$\pi/2<\Delta\phi<3\pi/2$.

The 2D and 3D collapse processes of Eqs.~(\ref{eqn:transv})
and~(\ref{eqn:3dcol}) occur on different time scales. It is likely
that both processes contributed to the $\sim90\%$ initial atom
loss
in the Strecker experiment; a measurement of the time
dependence of the atomic mass ejected from the trap during the
initial stages of soliton train formation may be able to determine
if one or both were dominant.  For the experimental parameters,
based on the collapse conditions and Fig.~\ref{fig:3}, primary
transverse collapse occurs on a time scale of $\,\lesssim 1.7$ ms,
primary 3D collapse occurs between 1.5 and 10~ms, and
secondary collapse occurs at $\,\gtrsim10$~ms.

In conclusion, the phenomenon of MI has been analyzed in the
case of a non-uniform initial state in the nonlinear Schr\"odinger equation
with a harmonic potential.  It was shown that linear density fringes, the
length scale of which depends both on space and time, seed the
nonlinear instability. Primary transverse collapse, primary
three-dimensional soliton collapse, and secondary
three-dimensional collapse due to soliton binary interactions were
discussed, and all were predicted to have played a role in the
Strecker experiment~\cite{strecker1}.

We thank Y.~Castin, R.~Hulet, and G.~Shlyapnikov for useful
discussions. L.~D.~Carr was supported by NSF grant no.~MPS-DRF
0104447. LKB is a unit of ENS and of Universit\'e Paris 6
associated to CNRS.  We thank the ECT$^{*}$ (Trento) and MPIPKS
(Dresden) for hosting us during various stages of this work.


\end{document}